# Performance Task using Video Analysis and Modelling to promote K12 eight practices of science

**Loo Kang Wee[1], Tze Kwang Leong[2]**
lawrence_WEE@moe.gov.sg, tzekwang.LEONG@rgs.edu.sg
[1]*Ministry of Education, Educational Technology Division, Singapore.*
[2]*Ministry of Education, Raffles Girl's School, Singapore.*

**Abstract**
We will share on the use of Tracker as a pedagogical tool in the effective learning and teaching of physics performance tasks taking root in some Singapore Grade 9 (Secondary 3) schools. We discuss the pedagogical use of Tracker help students to be like scientists in these 6 to 10 weeks where all Grade 9 students are to conduct a personal video analysis and where appropriate the 8 practices of sciences (1. ask question, 2. use models, 3. Plan and carry out investigation, 4. Analyse and interpret data, 5. Using mathematical and computational thinking, 6. Construct explanations, 7. Discuss from evidence and 8. Communicating information).
We will situate our sharing on actual students' work and discuss how tracker could be an effective pedagogical tool. Initial research findings suggest that allowing learners conduct performance task using Tracker, a free open source video analysis and modelling tool, guided by the 8 practices of sciences and engineering, could be an innovative and effective way to mentor authentic and meaningful learning.
Actual tracker files *.TRZ are down-able via Dropbox links on this links:
http://weelookang.blogspot.sg/2014/05/tracker-koaytzemin-student-video-roller.html
http://weelookang.blogspot.sg/2014/05/tracker-deeakdev-student-video.html
http://weelookang.blogspot.sg/2014/05/tracker-wangyuxing-student-video.html
http://weelookang.blogspot.sg/2014/05/tracker-dianielleteo-student-video.html
http://weelookang.blogspot.sg/2014/05/tracker-camelialim-student-video-ping.html

## What

We used the K12 science education framework[1] to guide our use of the video analysis[2] and modelling[3] approach to allow students to be like scientists.

## Why?

**Table 1**: Comparison of traditional lessons versus the scientist research

| *Traditional Lessons* | *Scientist Research* |
|---|---|
| Topics taught in isolation | Knowledge from various topics are required |
| Simplified theoretical scenario with many assumptions | Authentic collected data which often include anomaly |
| Knowledge apply to assessment questions | Knowledge apply to real world situations |
| Teacher directed; Teacher decide on question | Student directed; Ownership of research question |
| No differentiation; One size fits all | Research is individualized |

Table 1: comparison of traditional lessons versus the scientist research

Table 1 summarise the various reasons why high school science education should consider incorporating elements of the scientists' research as about a 10 week duration performance task as we speculate traditional lessons generally have weaknesses (Table 1 LEFT) in allowing students to experience[4] physics phenomena of the student's choice.

**How?**

Flip video

A series of YouTube instructional video were created by the authors to demonstrate simpler tasks such as a) installing Tracker b) generate *s* vs *t*, *v* vs *t* and *a* vs t graphs and c) analyse graphs by finding gradient[1] and area[2].

Classroom activities

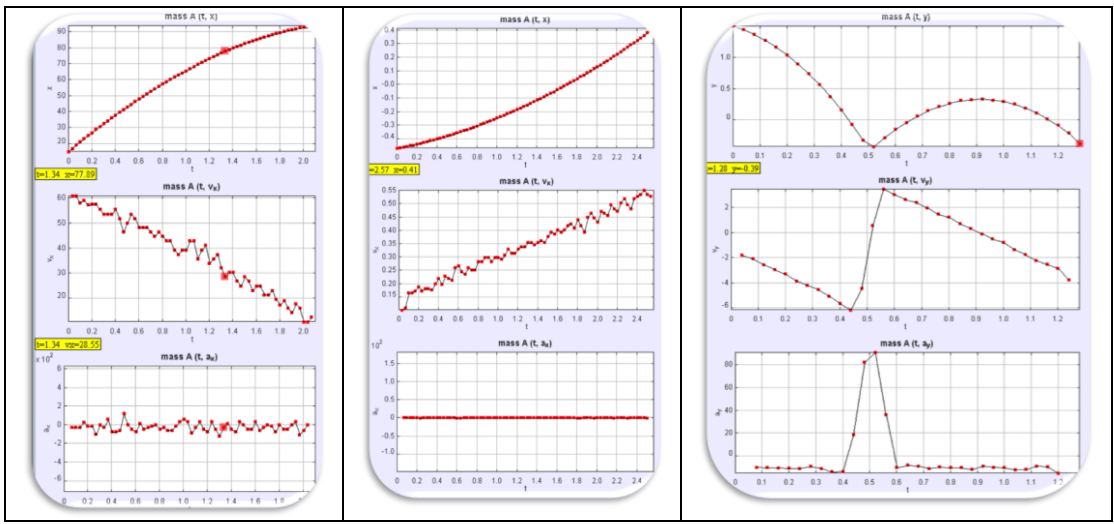

---

[1] https://www.youtube.com/watch?v=H_zrkl16BNs
[2] https://www.youtube.com/watch?v=7_TgOSMqRQs

Figure 1.  **LEFT to RIGHT** a) cart push on level slope slowing down at constant negative acceleration) b) cart on slope speeding up at constant acceleration and c) bouncing ball, complex accelerating motion divided into phases.

A set of 3 video analyses (Figure 1) were provided to the students in groups of 3-4, where they will discuss, present and critic on the other groups' presentation. As students were introduced to Tracker's interface in the flip video at their own free time earlier on, the lessons stand a higher chance of allowing richer discussions in class. The selected video are a) cart push on level slope slowing down at constant negative acceleration) b) cart on slope speeding up at constant acceleration and c) bouncing ball, complex accelerating motion divided into phases.

## Primer Scientist Activity

We also conducted a primer hands-on activity where a set of equipments (2 marbles, a ramp, metre rule etc) where students need to behave like a beginning scientists (Figure 2) such as
a. Practice 1: Define a problem
b. Practice 3: Plan out an investigation
c. Practice 4: Analyse and interpret the data
d. Practice 6: Construct explanations
Some interesting student's asking own questions involves experiments such as marble rolling up and down a ramp, bouncing marble down stairs or slope, projectile marble colliding with wall and colliding marbles[3].

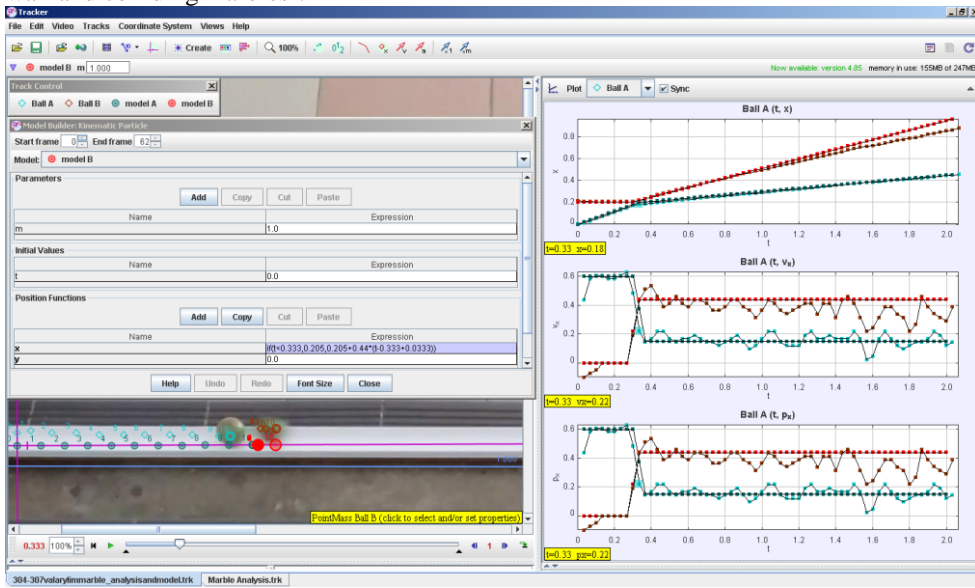

Figure 2.  Students directed video analysis inquiring on the inelastic collision of 2 marbles on a track. Teal trail is the right moving marble and the Red trail is the right moving marble. Model A and B were added on later by the authors for teacher professional network learning purposes.

---

[3] https://www.dropbox.com/s/4ypouk5hgc69hww/304-307valarylimmarble_analysisandmodel.trz

## Dynamics Topic integrative lesson from Kinematics

*Revisit gentle push on horizontal slope*

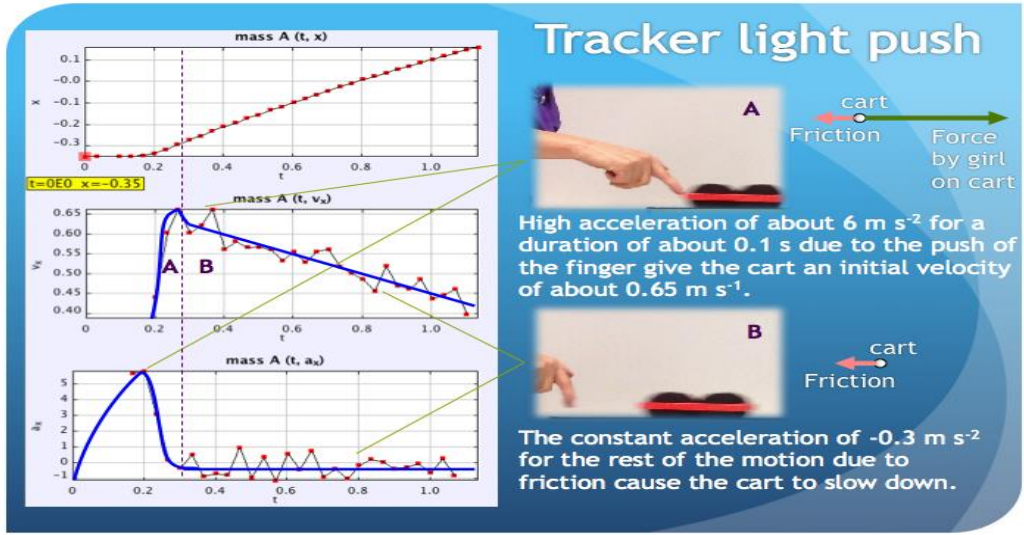

Figure 3.  A dynamics lesson using tracker again to illustrate frictional forces bridging from kinematics topics allows students to build understanding based on what they have already experience themselves in kinematics topics with the now more familiar tracker *x* vs *t*, *vx* vs *t* and *ax* vs *t* graphs.

Students were asked to suggest what forces cause the motion where all previous kinematics videos are revisited to discuss the dynamics (Figure 3). There was no enough time for teacher professional development[5] this year but the authors recommend getting students to use the Tracker's Dynamics model builder to incrementally suggest better models that represent the motion under investigations that has both practice 5: mathematical and computational thinking.

### *Practice 5: Mathematical thinking*

For example, to model frictional force motion (Figure 4) of the cart moving horizontally, students can mathematically determine the gradient of the *vx* vs *t* graph to determine acceleration in the *x* direction, *ax*.

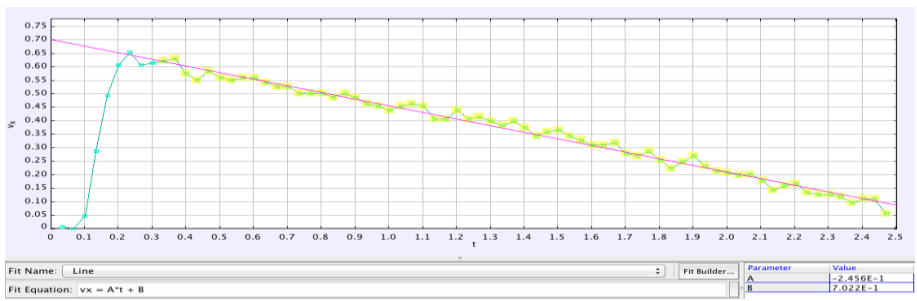

Figure 4. Tracker DataTool showing a Fit Line of $vx = -0.2456*t+0.7022$.

### Practice 5: Computational thinking[6]

Having determined $ax$ and knowing mass of cart to be $m = 0.2$ kg, the computational line (Figure 5) with the if statement allows the push force $fx = 3.2*m$ from $t = 0$ to 0.2 s, and frictional force $fx = -0.246*m$ to be present after time $t$ is greater than 0.2 s.

$$fx = if\,(t<0.2, 3.2*m, -0.246*m)$$

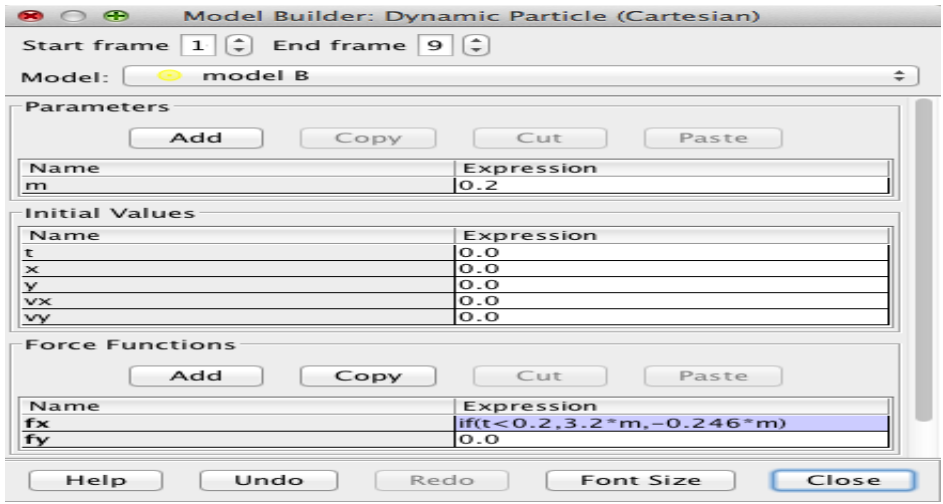

Figure 5. Tracker Dynamics Model Builder interface showing mass $m = 0.2$ kg and force model $fx = if\,(t<0.2, 3.2*m, -0.246*m)$.

*Atwood Machine*

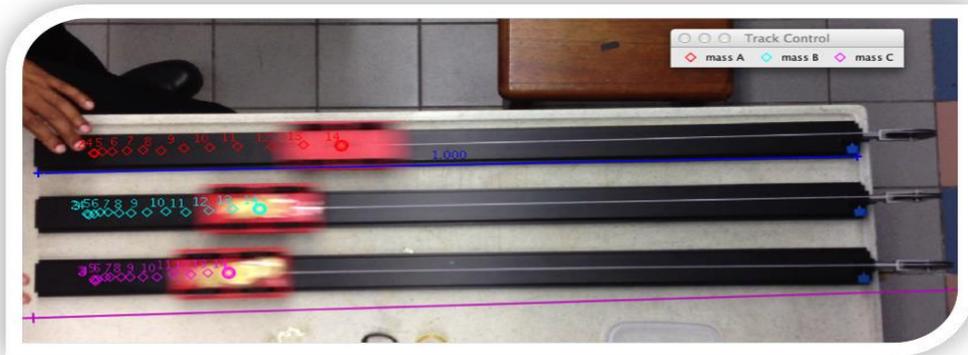

Figure 6.   Atwood Machine video showing 3 different mass, *m* carts, pulled by the same weight *F* (RIGHT) and their respectively accelerations *ax*.

Another example we tried was the Atwood machine (Figure 6) where 3 carts of different mass are pulled by identical weights. This activity aims to allow students to modelling instruction teaching approach to arrived at a social meaning making process that the resultant force *F* pulling the carts is related to the carts' acceleration *a* and the mass of the carts as in Newton's second law.

$$F = ma$$

## Assessment of Performance Task

*Meaningful Scenario*

We gave the students the scenario that mimics real scientists. For example, "You are a scientist who is tasked by A*Star (A local Research initiative) to explain a complex motion and the cause of the motion. You are to record a video of a moving object and to analyze the kinematics and dynamics involved in the motion with the aid of Tracker software. Your report will help the scientific community better understand the complex motion."

*Assessment Rubrics(20% of exam score)*

| ASSESSMENT RUBRICS ON PT ON ANALYSIS OF MOTION USING VIDEO TRACKER YEAR THREE (RP PHYSICS) | | | | | |
|---|---|---|---|---|---|
| **Level**<br><br>**Criteria** | **Excellent**<br><br>**(4m)** | **Proficient**<br><br>**(3m)** | **Adequate**<br><br>**(2m)** | **Limited**<br><br>**(1m)** | **Insufficient / No evidence**<br><br>**(0m)** |
| **Identify motion to be investigated (C1)** | Identify a motion that is **complex** and **well defined** and involve non-rigid body, multiple objects or multiple phases | Identify a **complex** motion that is **well defined**. | Identify a motion that is **well defined**. | Identify a motion that is **ill-defined**. | |
| **Plan the procedure and filming (C2)** | Use a **comprehensive** and **detailed** procedure to film object in order to ensure **precision** and **accuracy** of measurement. | Use a **clear** and **workable** procedure to film object in order to ensure **precision** and **accuracy** of measurement. | Use a **simplistic** procedure to film object with **some** consideration of **accuracy** of measurement. | Use an **ambiguous** procedure to film object with **little** consideration of **accuracy** of measurement. | |
| **Present graphs with annotation (C3)** | Present graphs **logically** and **clearly** in an appropriate form with relevant **annotation**. | Present graphs **reasonably well** in an appropriate form with relevant **annotation**. | Present relevant but **incomplete** set of graphs | Present graphs with **severe conceptual error**. | |
| **Provide a discussion of the motion (C4)** | Provide a **detailed** and **comprehensive discussion** of the motion. | Provide a **relevant discussion** of the motion with **no errors**. | Provide a discussion of motion with **some minor errors**. | Provide a discussion of motion with **severe conceptual errors**. | |
| **Explain the forces in relation to the motion (C5)** | Explain the forces in relation to the motion by **consistently** making **accurate** inferences from graphs and video. | Explain the forces in relation to the motion by making **some accurate** inferences from graphs | Explain the forces in relation to the motion by making inferences from graphs with **some errors**. | Explain the forces in relation to the motion by with **little attempt** to make inferences from graphs. | |

Table 2: Assessment Rubrics communicating clearly the expected performance indicators of excellent task

We found a constantly reference to assessment rubrics, see Table 2, to be able to provide the extrinsic motivation to get students to perform their own performance video tasks.

Close Teacher Mentorship

For the learning to go well, we find that a weekly schedule consultation time-table to be useful as students are given opportunities to discuss their analysis with the teachers both in class and outside class. Suggestions were made by teachers to refine their video, given direction for further readings or suggestion on analysis.

Students' Pre-Post Perception Survey

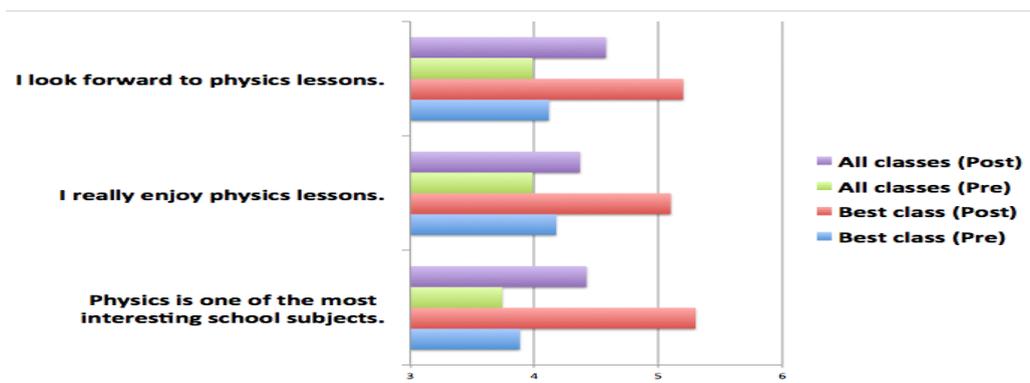

Figure 7.  **(N=273 pre-post)** Students' perception survey on a Likert scale from 1 (strongly disagree) to 6 (strongly agree), middle is 3.5 point

Our initial analysis of N=273 students' pre-post perception survey on the experience of the performance task suggests only the best class registered a significant positive change (Figure 7) in the affective domains like "I look forward to physics lessons", "I really enjoy physics lessons" and "Physics is one of the most interesting school subjects" etc while the 273 students in the Secondary three level had smaller positive changes in the self reporting perfection survey.
More sample tracker files *.TRZ are down-able via Dropbox links on this links:
http://weelookang.blogspot.sg/2014/05/tracker-koaytzemin-student-video-roller.html
http://weelookang.blogspot.sg/2014/05/tracker-deeakdev-student-video.html
http://weelookang.blogspot.sg/2014/05/tracker-wangyuxing-student-video.html
http://weelookang.blogspot.sg/2014/05/tracker-dianielleteo-student-video.html
http://weelookang.blogspot.sg/2014/05/tracker-camelialim-student-video-ping.html

**Conclusion**

This paper describes the what, why and how we promoted our students to be more like scientists in the context of our Secondary Three Physics performance task using Tracker and K12 Science Framework to guide our approach. The creative commons attribution resources we have created through our mentorship and lesson packages developed, forms the Singapore Tracker Digital Library collection downloadable through Tracker as a Shared Library (Figure 8) as well as the following URL http://iwant2study.org/lookangejss/ for the benefit of all.

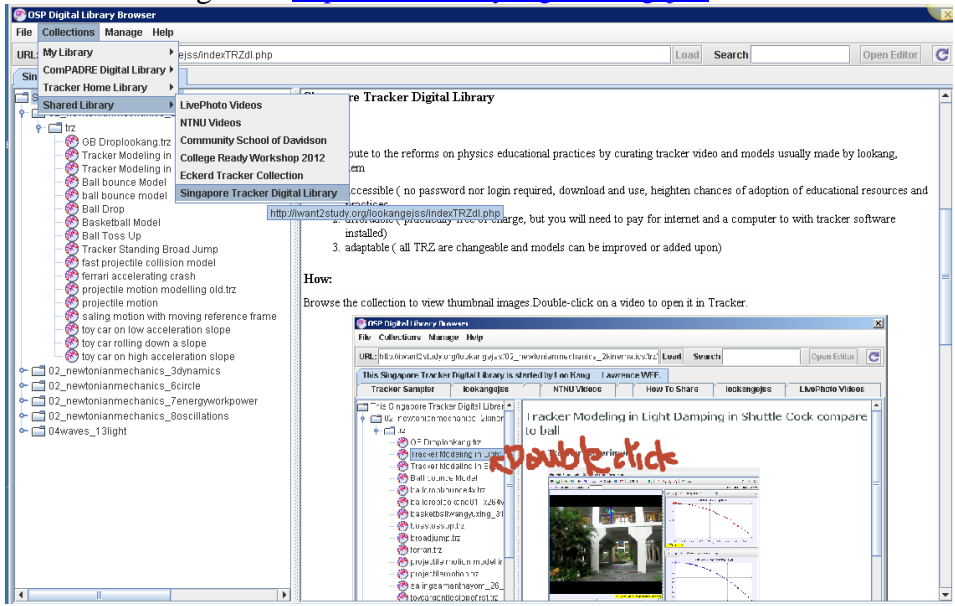

Figure 8. Officially in Tracker as Shared Library **http://iwant2study.org/lookangejss/indexTRZdl.php**

**Acknowledgement**

We wish to acknowledge the passionate contributions of Douglas Brown, Wolfgang Christian, Mario Belloni, Anne Cox, Francisco Esquembre, Harvey Gould, Bill Junkin, Aaron Titus and Jan Tobochnik for their creation of Tracker video analysis and modeling tool.

This research is made possible; thanks to the eduLab project NRF2013-EDU001-EL017 Becoming Scientists through Video Analysis, awarded by the National Research Foundation, Singapore in collaboration with National Institute of Education, Singapore and the Ministry of Education (MOE), Singapore.

Loo Kang Lawrence WEE
Ministry of Education, Educational Technology Division, Singapore.
1 North Buona Vista Drive, Singapore 138675
Singapore
e-mail: lawrence_wee@moe.gov.sg